**Epitaxial PbGeSe thin films and their photoluminescence in the mid-wave infrared**

Kelly Xiao[1], Bryce Wong[1], Jarod Meyer[1], Leland Nordin[2,3], Kunal Mukherjee[1*]

[1]Department of Materials Science and Engineering, Stanford University, Stanford, CA 94305, USA

[2]Department of Materials Science and Engineering, University of Central Florida, Orlando, FL 32816, USA

[3]CREOL, The College of Optics and Photonics, University of Central Florida, Orlando, FL 32816, USA

## ABSTRACT

PbSe is a narrow bandgap IV-VI compound semiconductor with application in mid-wave infrared optoelectronics, thermoelectrics, and quantum devices. Alkaline earth or rare earth elements such as Sr and Eu can substitute Pb to widen the bandgap of PbSe in heterostructure devices, but they come with challenges such as deteriorating optical and electronic properties, even in dilute concentrations due to their dissimilar atomic nature. We substitute Pb instead with column-IV Ge and assess the potential of rocksalt phase PbGeSe as a wider bandgap semiconductor in thin films grown by molecular beam epitaxy on GaAs substrates. Low sticking of GeSe adatoms requires synthesis temperatures below 260 °C to incorporate Ge, but this yields poor structural and compositional uniformity as determined by X-ray diffraction. Consequently, as-grown films in the range $Pb_{0.94}Ge_{0.06}Se$ to $Pb_{0.83}Ge_{0.17}Se$ (6–17% Ge) show much less bandgap widening in photoluminescence than prior work on bulk crystals using absorption. We observe that post-growth rapid thermal annealing at temperatures of 375–450 °C improves the crystal quality and recovers bandgap widening. Rapid interdiffusion of Ge during annealing, however, remains a challenge in harnessing such PbGeSe materials for compositionally sharp heterostructures. Annealed 17% Ge films emit light at 3–3.1 µm with minimal shift in wavelength versus temperature. These samples are wider in bandgap than PbSe films by 55 meV at room temperature and the widening increases to 160 meV at 80 K, thanks to sharply different dependence of bandgap on temperature in PbSe and PbGeSe.

* Author to whom correspondence should be addressed: kunalm@stanford.edu

## I. INTRODUCTION

The narrow bandgap rocksalt IV-VI semiconductors, exemplified by PbSe and PbTe, are well known in mid-wave and long-wave infrared (MWIR: 3-5 μm, and LWIR: 8-14 μm) optoelectronics,[1,2] intermediate temperature thermoelectrics,[3] and most recently in quantum information sciences. These narrow bandgaps are situated in the molecular fingerprint region and are therefore particularly well-suited for gas sensing, with recent demonstrations of methane detection at 3.3 μm from band-engineered PbSe devices.[4] Focusing on optoelectronics, wider bandgap alloys in the same family would enable PbSe-based devices to expand to the short-wave infrared (SWIR: 0.7-2.5 μm) while retaining the attractive properties of IV-VI materials, including low Auger recombination and potential defect tolerance.[5,6] Wider bandgap alloys as barrier materials also improve the operating characteristics of MWIR and LWIR devices through increased carrier confinement in type-I quantum wells and double heterostructures.

To date, popular wide-gap barrier materials for PbSe have been alkaline-earth and rare-earth alloys such as PbSrSe and PbEuSe, respectively. Only a few percent of Sr and Eu substituting Pb is sufficient to widen the bandgap (~30 meV/%-Sr or %-Eu), enabling multi-quantum well structures to be grown with a small lattice-mismatch.[7] Substitution by very dissimilar or so-called highly mismatched elements[8] (even if isovalent), however, can be challenging due to solubility issues[9,10] and may even diminish properties of both bulk semiconductors and heterostructures. Sr and Eu are indeed highly dissimilar to Pb, and it has been reported that even 1–2% Sr or Eu alloying leads to a 10–100× reduction in photoluminescence (PL) intensity and Hall mobility, even at cryogenic temperatures.[2,11,12] Besides alkaline earth and rare-earths, some cations such as Cd (up to 10% in polycrystalline films) has also been reported to widen bandgap;[4] however, effects on PL intensity or other (opto)electronic properties remain unclarified. Substitution of the Pb site with similar column IV elements (Si, Ge, Sn) within the rocksalt phase remains of interest for bandgap tuning.

Ordinarily, substitution with a lighter isovalent element increases the bandgap, but substituting heavy Pb with Sn anomalously reduces the bandgap.[13] This is fortuitous for long wavelength IV-VI devices, but the need for a wider bandgap material remains. Ge is lighter than Pb and conventionally increases the bandgap upon substitution.[14] Simple junction devices with PbGeTe confirm bandgap widening with Ge alloying,[15] but PbGeTe undergoes a structural transition to a ferroelectric phase at cryogenic temperatures for even dilute Ge concentrations, which is undesirable for cooled infrared devices.[16,17] Much less is known about the properties of PbGeSe alloys, with only a handful of studies on bulk samples and none on epitaxial thin films to our knowledge.[14,18–20] Krebs et al. synthesized bulk samples of PbGeSe and found a decreasing lattice parameter up to a solubility limit of 9% Ge (notated as group-IV site percent throughout this paper) in the rocksalt phase at 460–480 °C.[18] The low solubility was validated in recent work on PbGeSe thermoelectrics with a maximum 50–60 meV blueshift in the optical absorption edge at 6-9% Ge.[20] On the other hand, Nikolic reports up to 40% Ge solubility in bulk samples and notes a linear blueshift in the absorption edge with increasing Ge. The bandgap widened from 0.26 eV of PbSe to 0.4 eV for $Pb_{0.6}Ge_{0.4}Se$, yielding a 140 meV confinement at room temperature adequate for many devices.[14] These bulk samples were synthesized at a higher temperature of 630 °C and quenched to room temperature, potentially accounting for increased solubility of Ge. In this work, we revisit PbGeSe to assess its feasibility as a wider bandgap semiconductor to PbSe. We present results on the epitaxial growth of PbGeSe films on GaAs templates by molecular beam epitaxy and use PL to demonstrate properties of this alloy relevant to infrared optoelectronics.

## II. METHODS

Films were synthesized using a Riber Compact 21 IV-VI system equipped with compound PbSe and GeSe dual zone effusion cells. Compound sources are used in IV-VI growth (in contrast with elemental sources used in conventional III-V growth) to improve stoichiometry control.[11,12,21–23] Arsenic-capped homoepitaxial GaAs (001) substrates were prepared *ex-situ* in a Veeco Gen III chamber, indium-bonded to Mo platens, and loaded into the IV-VI chamber. The amorphous As capping layer was thermally desorbed to expose the preserved homoepitaxial GaAs (001) template before initiating growth. To promote higher quality growth of PbGeSe, a 12 nm epitaxial PbSe buffer layer was first grown on GaAs using a nucleation sequence discussed in Haidet *et al*.[24]

Optimal growth temperatures for the main PbGeSe layer were initially investigated in Samples T1-T4 by holding constant the PbSe beam equivalent pressure (BEP) at $3 \times 10^{-7}$ Torr (corresponding to a growth rate of 0.42 Å/s of PbSe) and the GeSe BEP at $9.8\text{-}10 \times 10^{-8}$ Torr. Then, to generate a series of Ge compositions, we sweep two growth parameters of tuning relative fluxes through GeSe BEP (Samples P1-P3) and growth rate (Samples R1-R3). The growth conditions of the aforementioned samples are accordingly summarized in Table I. Higher growth rates of PbGeSe were achieved by scaling both the PbSe and GeSe BEPs by identical factors greater than unity. The targeted thickness for the PbGeSe films was ~130-150 nm. For post-growth rapid thermal annealing (RTA) treatments, films were capped with approximately 36 nm of chemical vapor deposited $SiO_2$ and thermally annealed under an inert $N_2$ atmosphere in an AllWin 610 RTA.

Samples were characterized using photoluminescence (PL) and X-ray diffraction (XRD). For quasi-continuous wave PL measurements, samples were pumped using a 1 W, 808 nm laser that was modulated at 10 kHz with a 50% duty cycle square wave. PL spectra were resolved using a Bruker Invenio-R Fourier-transform infrared spectrometer operating in a step-scan mode. Temperature-dependent PL measurements were taken in a liquid-nitrogen cooled cryostat with a $CaF_2$ window. Cubic (224) reflection XRD reciprocal space maps (RSMs) were collected on a PANanalytical Empyrean diffractometer using Cu-K$\alpha_1$ radiation, mapped in grazing exit geometry. Film compositions were estimated from RSMs assuming biaxial strain and using the linear relationship between Ge-% and lattice constant obtained by Nikolic in bulk PbGeSe.[14]

**Table I.** Sample IDs with growth temperature ($T_g$), GeSe and PbSe BEPs, and estimated Ge composition.

| Sample | $T_g$ (°C) | GeSe BEP (Torr) | PbSe BEP (Torr) | Estimated Ge (%) |
|---|---|---|---|---|
| **T1** | 195 | same sample as P3, refer to P3 details | | |
| **T2** | 210 | same sample as R1, refer to R1 details | | |
| **T3** | 230 | $1.0 \times 10^{-7}$ | $3.0 \times 10^{-7}$ | -- |
| **T4** | 260 | $1.0 \times 10^{-7}$ | $3.0 \times 10^{-7}$ | -- |
| **PbSe** | 195 | -- | $3.0 \times 10^{-7}$ | 0 |
| **P1** | 195 | $4.6 \times 10^{-8}$ | $3.0 \times 10^{-7}$ | 6 |
| **P2** | 195 | $7.0 \times 10^{-8}$ | $3.0 \times 10^{-7}$ | 8 |
| **P3 / T1** | 195 | $9.8 \times 10^{-8}$ | $3.0 \times 10^{-7}$ | 11 |
| **R1 / T2** | 210 | $1.0 \times 10^{-7}$ | $3.0 \times 10^{-7}$ | 7 |
| **R2** | 210 | $2.0 \times 10^{-7}$ | $6.0 \times 10^{-7}$ | 13 |
| **R3** | 210 | $4.0 \times 10^{-7}$ | $1.2 \times 10^{-6}$ | 17 |

## III. RESULTS AND DISCUSSION

### A. Identifying a growth window to incorporate Ge

Figure 1a shows a schematic of the PbSe buffer and PbGeSe layer of all the samples studied in this paper. Differences in the thermophysics of GeSe and PbSe make temperature an important parameter in the synthesis of PbGeSe alloys. We have previously deposited PbSe up to 340 °C by molecular beam epitaxy (MBE) with unity sticking of adatoms, but near this temperature, GeSe exhibits orders of magnitude higher sublimation vapor pressure compared to PbSe.[25,26] Therefore, we expect PbGeSe alloys require lower growth temperatures for incorporation. Very low growth temperatures may, however, result in amorphous films or domains as GeSe is also a glass former unlike PbSe.[27] Considering these two extremes, we investigate the degree of incorporation of Ge into PbSe using a series of intermediate growth temperatures between 195–260 °C while holding GeSe and PbSe BEPs constant at $9.8\text{-}10\times10^{-8}$ Torr and $3\times10^{-7}$ Torr, respectively (Samples T1-T4).

Reflection high energy electron diffraction (RHEED) patterns for Samples T1-T3 at the end of growth are shown in Figure 1b. PbGeSe RHEED begins to exhibit increasing spottiness along the $(1 \times 1)$ vertical streaks with decreasing growth temperature, potentially associated with surface roughening. Figure 1c and 1d show the PL spectrum and bandgap estimated using the maximum of the PL energy derivative, a method reported in Webster et al.[28] We use the first derivative method in lieu of absorption due to the very thin films grown, and note that this PL method is sensitive to the onset energy at which the optical joint density of states increases most

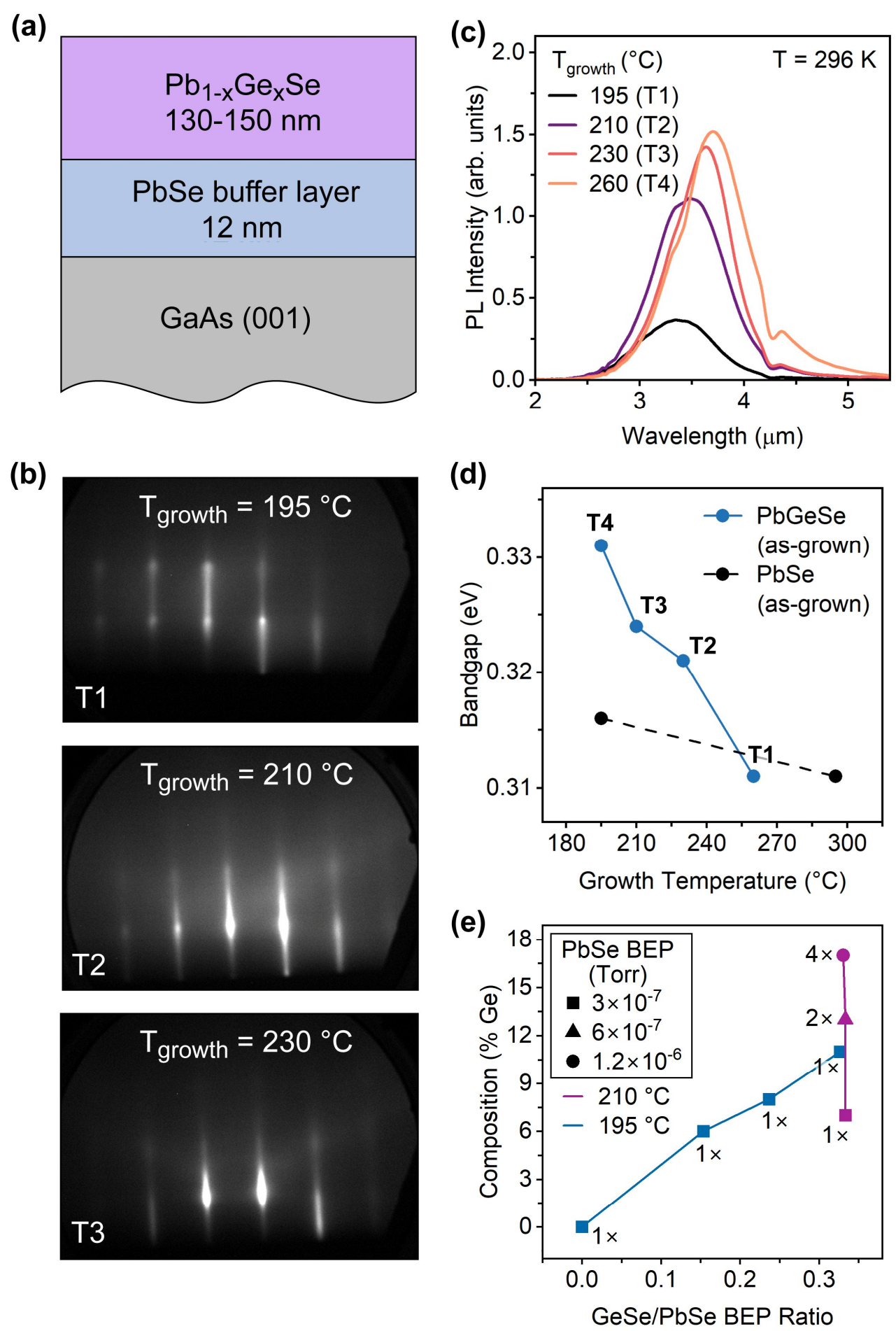

**Figure 1.** (a) Schematic diagram of grown layer structure on GaAs (001). (b) RHEED images of PbGeSe film at growth temperatures of 195, 210, and 230 °C (Samples T1-T3). (c) PL spectra of PbGeSe films synthesized between 195–265 °C (Samples T1-T4). Intensity and peak emission wavelength are dependent on growth temperature (and the related sticking coefficients). The consistent dip at 4.3 μm is due to atmospheric $CO_2$ absorption. (d) Extracted bandgaps for Samples T1-T4 compared to 195 °C PbSe and an additional 295 °C PbSe. (e) Measured GeSe/PbSe BEP ratios and the estimated compositions for films grown by sweeping GeSe BEP (for a fixed PbSe BEP) at 195 °C and sweeping growth rate (for a fixed GeSe/PbSe ratio) at 210 °C. The multiplier beside each point represents scaling of the base PbSe growth rate of 0.42 Å/s.

rapidly—the bandgap in ideal semiconductors. In nonideal semiconductors with inhomogeneous band edges due to localization effects or compositional fluctuations,[29,30] the first derivative method might show only the narrowest gap with significant optical density of states. This information is, nevertheless, just as relevant for many applications as a proxy for the bandgap. At our highest growth temperature of 260 °C, the PbGeSe bandgap approaches that of a reference PbSe, which we attribute to complete Ge re-evaporation. We show in Figure 1d the bandgap of PbSe at two growth temperatures to highlight the small but real effect of thermal expansion strain from the GaAs substrate.

Importantly, degradation in RHEED with decreasing growth temperatures appears alongside a monotonous blue shift in emission wavelength (and decreasing PL intensity) in Figure 1c. This is indicative that the films have been grown entering a GeSe sticking regime and that the bandgap and surface roughness of the alloy are increasing due to Ge incorporation. Lowering the growth temperature while keeping constituent fluxes constant was recently also shown to increase Ge content in PbGeTe alloys by MBE.[17]

Ideally, the edge of the unity sticking regime of Ge is important for robust high-quality growths—this would have been observed as the beginning of a plateau in bandgap on the lower range of growth temperatures. Based on Figure 1c, we prioritize PL intensity and concerns of amorphization and therefore utilize the 195–210 °C range for subsequent growths. Despite no conclusive evidence of unity Ge sticking, we demonstrate control over the %-Ge composition by varying GeSe BEPs for a fixed PbSe BEP of $3.0 \times 10^{-7}$ Torr at a 195 °C growth temperature (Samples P1-P3). Figure 1e shows a nearly linear trend in XRD-derived PbGeSe compositions up to 11% Ge plotted against BEP ratios (XRD data is shown in Figure 4 and discussed separately). The 195 °C PbSe reference is displayed as part of this trend. Also on Figure 1e, we show that for additional samples grown at 210 °C, scaling both the PbSe and GeSe BEPs by a factor greater than unity (i.e. a growth rate increase) is an alternative method to increase the Ge composition (Samples R1-R3). Scaling the BEPs by factors of 2 and 4 to achieve growth rate increases of nearly 2× and 4×, respectively, yields 13% and 17% Ge, both metastable alloy compositions. This increase in Ge content with growth rate is a symptom of the non-unity sticking regime and points to the kinetics of Ge re-evaporation that may be slowed down by burying GeSe (or Ge) adatoms with PbSe. The films in Figure 1e, which span six Ge compositions from 6-17%, are the focus of the remaining discussion.

### B. Post-growth thermal annealing

Notable features of the as-grown PbGeSe films included dimmed PL intensity compared to the binary, and perhaps more unexpectedly, the incomplete extent of blue shifting in the peak emission wavelength with increasing Ge as Nikolic reports via absorption measurements in bulk PbGeSe.[14] These phenomena will be discussed shortly in conjunction with Figure 3. With the goal of first recovering photoluminescence intensity in PbGeSe, we began by thermally annealing the highest composition film ($X_{Ge} = 17\%$, Sample R3) as a test case. This annealing process is partially motivated by our previous work where thin films of PbSe and PbSnSe alloys capped with $SiO_2$ exhibited improved luminescence after annealing.[31,32] Figure 2 showcases a remarkable improvement in PbGeSe PL spectra after annealing up to 450 °C, especially in terms of blue shifting of the peak emission wavelength. The PL spectra is shown after individual annealing steps taken to optimize the RTA temperature for the 17% Ge film. The PL spectrum evolves with cumulative and progressively increasing annealing temperatures (60 s anneal time, each). Raising the annealing temperature to 375 °C causes a significant blueshift in peak emission wavelength from 3.3 μm to ~3 μm. Further increasing temperature to 450 °C does not result in additional blueshift but increases PL emission intensity. Increasing the annealing time at 450 °C to 120 s and 240 s did not improve the PL spectra (not shown).

Annealing at 450 °C for 60 s is similarly effective on other compositions. Figure 3 compares the as-grown and post-annealed PL spectra from the six total PbGeSe compositions from 6% to 17% Ge discussed above, including the PbSe reference. The PL spectra for 17% Ge has been remeasured following a direct, and not cumulative, annealing step to 450 °C to be directly comparable. As alluded previously, the as-grown material demonstrates an unexpected stalling or non-monotonic behavior of the peak emission wavelength. Figure 3a captures the stagnant emission from films in the GeSe BEP series for compositions up to 11% (Samples P1-P3). In contrast, their spectra taken after annealing in Figure 3b successfully show increasing blueshifts with composition. Likewise, comparing Figures 3c and 3d, as-grown spectra from PbGeSe grown at varied rates also show increasing blueshifts upon annealing. In Figure 3e, we see the blueshifted spectra are simultaneously characterized by narrower PL peak widths across all compositions, suggesting partial reduction in disorder and band edge inhomogeneity occurs at the elevated anneal temperatures. The PbGeSe PL FWHMs do not surpass that of the PbSe baseline in both as-grown and annealed conditions.

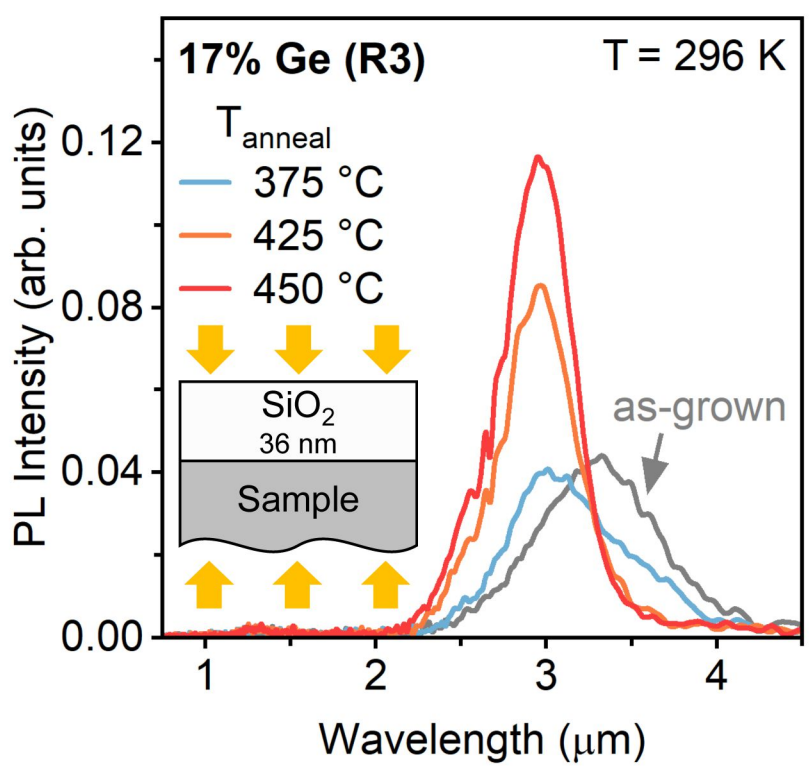

**Figure 2.** Photoluminescence spectra of 17% Ge film (Sample R3) following annealing at 375 °C to 450 °C for 60 seconds, with the spectrum for the as-grown sample shown in grey. A shift in wavelength is observed at 375 °C and intensity gains are observed up to 450 °C. The inset schematic depicts the $SiO_2$-capped annealed structure.

The stalled versus moving blueshifts in peak wavelength for as-grown versus annealed PbGeSe are reflected in the estimated bandgap trends within Figure 3f. From GeSe BEP sweeping, incorporation of 6% Ge (Sample P1) produces a bandgap widening of about 15 meV relative to as-grown PbSe. Further incorporation of 11% Ge (Sample P3) results in comparatively little widening of only 18 meV. Upon annealing, the bandgap exhibits a more linear dependence on Ge content. Incorporation of 6% Ge produces a widening of about 21 meV relative to annealed PbSe, and 11% Ge produces a larger net widening of approximately 42 meV at room temperature (approximately 4 meV/%-Ge). We find that the slope of the bandgap vs. Ge composition agrees closely with data from absorption measurements in bulk PbGeSe crystals by Nikolic,[14] and note that much of the sizeable offset to our data (even for PbSe) arises due to thermal expansion mismatch strain from the GaAs substrate.[33]

In Figure 3f, next looking at the films grown at higher rates (Samples R1-R3), a more severe plateau in bandgap is apparent in the as-grown samples—compositions spanning 7–17% result in only a net widening of roughly 25 meV. Again, annealing increases the bandgap; however, the slope of bandgap vs. composition now slightly deviates from linearity. This hints that the growth rate parameter and GeSe flux parameter are not equivalent Ge incorporation mechanisms, and that we have yet to achieve systematic control over the effects of growth rate. Nevertheless, the highest composition of 17% yields a net widening of 55 meV relative to annealed PbSe (~3 meV/%-Ge). This value is already ~2$k_BT$ at room temperature, where $k_B$ is Boltzmann's constant, and indicates good potential for PbSe/PbGeSe heterostructures once band alignments are also known. The rocksalt GeSe endpoint bandgap or maximum potential confinement is difficult to predict, as the rocksalt phase of GeSe is metastable and lacks optical data.[34,35] The stable phase of GeSe is a layered distorted rocksalt, and the composition threshold for metastable epitaxial rocksalt GeSe not been assessed in this study. Viewing Figures 3a-d together, we see that alloying with Ge nevertheless reduces PL intensity at higher Ge compositions even with annealing—that is, the bandgap widening comes the expense of reduced emission intensity at room temperature. Although not as deleterious as Eu or Sr, future work on the upper limit of Ge incorporation in the rocksalt phase thin films should consider ways to avoid quenching of the PL intensity.

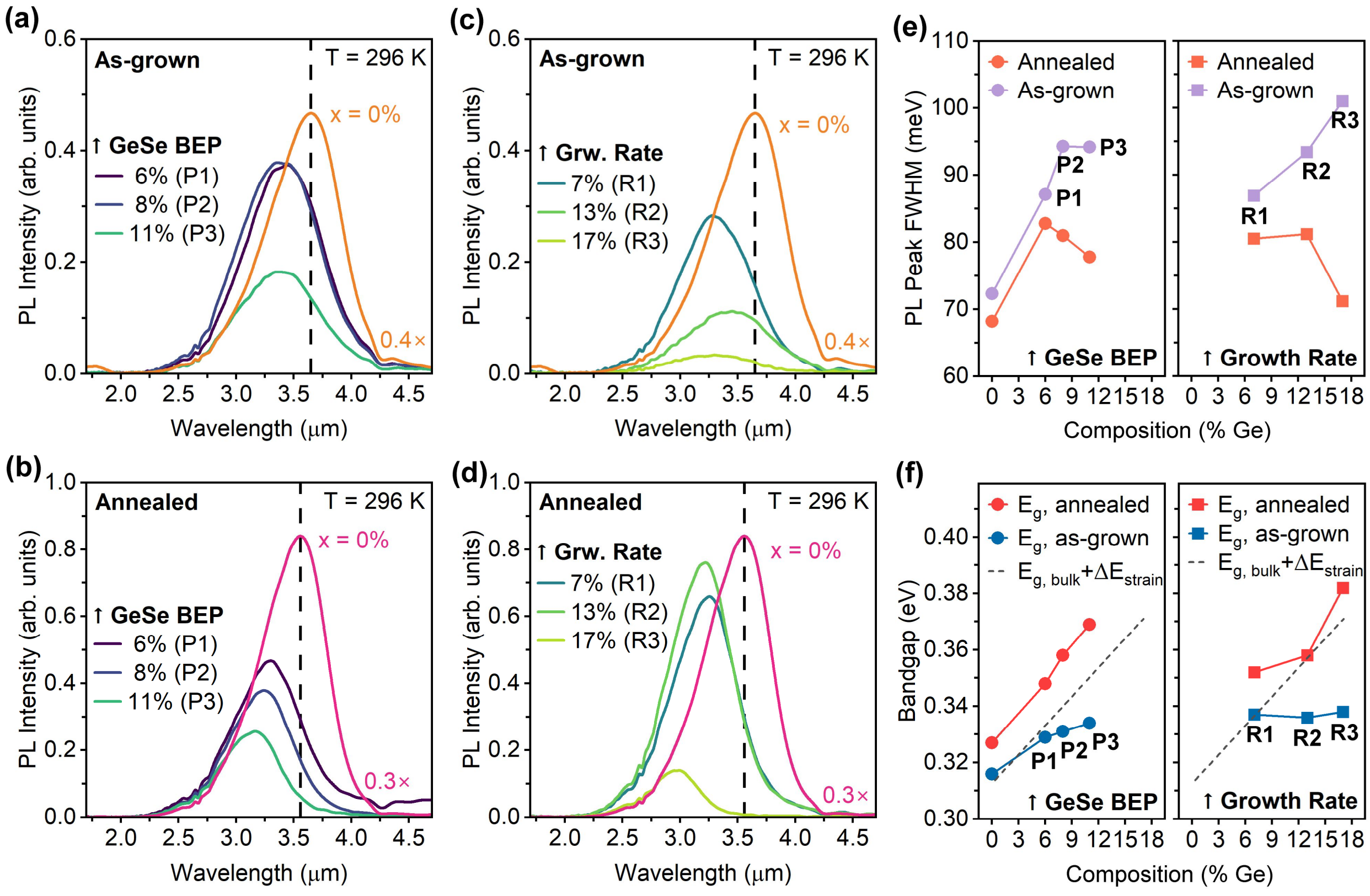

**Figure 3.** (a) Room temperature photoluminescence for as-grown 6-11% PbGeSe films grown with varying GeSe BEPs (Samples P1-P3). The spectrum for as-grown 195 °C PbSe scaled by 0.4× is plotted for reference. (b) PL spectra for Samples P1-P3 and PbSe reference, annealed at 450 °C for 60 s. The annealed PbSe spectra is scaled by 0.3×. (c) Room temperature PL for as-grown 7-17% PbGeSe films grown under varying growth rates (Samples R1-R3). The spectrum for as-grown 195 °C PbSe scaled by 0.4× is plotted for reference. (d) PL spectra for Samples R1-R3 and PbSe reference, annealed at 450 °C for 60 s. The annealed PbSe spectra is scaled by 0.3×. The dip in PL spectra at 4.3 μm is due to atmospheric $CO_2$ absorption. (e) Full-width half maximum of PL peaks vs. Ge composition for PbSe, Samples P1-P3 and R1-R3, as-grown and annealed. (f) Extracted bandgap vs. Ge composition for PbSe, Samples P1-P3 and R1-R3, as-grown and annealed. The bandgap of bulk PbGeSe with additional calculated blueshift from thermal strain (~45 meV) is indicated by the dashed line. Bulk data is retrieved from Nikolic (Ref. 14).

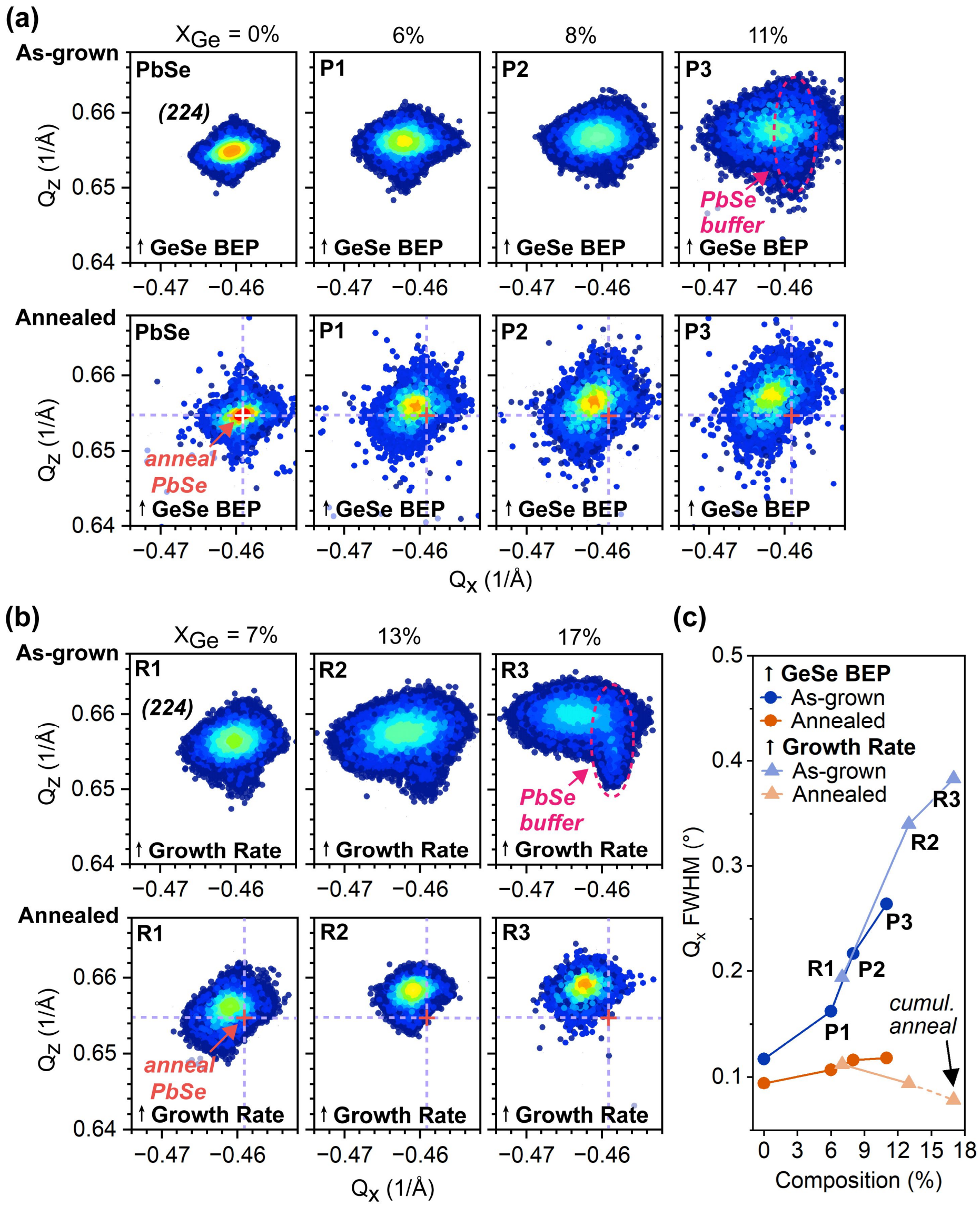

**Figure 4.** Reciprocal space maps of the PbGeSe (224) reflection prior to and after 450 °C annealing for (a) compositions achieved in the GeSe BEP series (Samples P1-P3) and (b) compositions achieved in the growth rate series (Samples R1-R3). The PbSe buffer (224) reflection is highlighted by the dashed oval for as-grown Samples P3. The annealed PbSe $Q_x$/$Q_z$ coordinates are also shown for reference. (c) FWHM trends of the PbGeSe (224) reflection along $Q_x$ for as-grown and annealed films. Only the 17% film (Sample R3) has undergone cumulative anneals.

We analyze the effects of annealing on the film structure in more detail using XRD reciprocal space maps (RSMs). Figures 4a and 4b show the PbGeSe (224) RSMs for the as-grown versus annealed conditions, across samples controlling the GeSe BEP and growth rate. First, we discuss the features of the as-grown film RSMs in both series. The as-grown films exhibit a higher intensity (224) component corresponding to the main PbGeSe layer, and another lower intensity (224) component elongated along $Q_z$ corresponding to the thin 12 nm PbSe buffer layer. These two components overlap for the lower compositions but become more distinct for higher compositions, such as in as-grown 17% Ge where the PbSe buffer resembles an arm extending from the primary PbGeSe peak (see as-grown Sample R3 in Figure 4b). We attribute the increasing separation between the components to a change in lattice dimensions with Ge alloying. Additionally, the PbGeSe (224) reflection is noticeably broad and diffuse primarily in the lateral $Q_x$ direction as greater percentages of Ge are incorporated. Bearing characteristics of diffraction dominated by coherence length effects, the reflections suggest Ge alloying degrades the film microstructure to yield reduced correlation lengths.

Annealing induces a significant transformation in peak shape across all samples. The PbSe buffer (224) reflection is now absent and suggests that the thin 12 nm PbSe buffer and much thicker PbGeSe film have intermixed. The final coalesced peaks have sharpened and adopt a shape dominated by mosaic broadening (in the ω-direction) rather than coherence length. The well-defined PbGeSe peak centers are clearly separated from the annealed PbSe reference and their reciprocal space coordinates indicate Ge substitution on the Pb sites yields shrunken rocksalt lattice dimensions. A decreasing lattice constant is in accordance with lattice constant trends in bulk PbGeSe[14,18], information we use to obtain the Ge compositions in Table I. In Figure 4c, the (224) angular spread along the $Q_x$ direction has been extracted to qualitatively capture trends in lateral correlation for all Ge compositions. Annealing yields dramatic structural improvement, suggested by the lower $Q_x$ widths which all fall to comparable values. Note that the slightly lower broadening for the 17% Ge sample is due to cumulative annealing (see Figure 2) and indicates some room for further structural improvement in all the other Ge-alloyed samples. We find the (224) in-plane spread is only marginally improved for the pure PbSe subjected to a 450 °C anneal. Thus, annealing is particularly necessary to improve the film quality of PbGeSe as-grown alloys.

### C. Temperature dependence of photoluminescence

We finally look at the temperature dependent trends in the PL intensity and bandgap. Figures 5a-b, 5c-d, and 5e-f show the 80–296 K PL spectra of the as-grown and same annealed film for compositions of 6%, 11%, and 17% Ge, respectively. Shifting of the spectra towards lower wavelengths and narrowing peak widths occurs with the higher compositions. We have also measured the temperature-dependent spectra of as-grown PbSe (Figure S1).

Figure 6a summarizes trends in bandgap with temperature, omitting a few points for when the atmospheric $CO_2$ absorption line (4.3 μm) modulates the peak shape and makes estimating bandgap challenging. We note both as-grown and annealed 6% Ge spectra have strong $CO_2$ absorption features and therefore this sample is omitted entirely from Figure 6a. Although opposite to conventional semiconductors, we find the bandgap of the PbSe film narrows with decreasing temperature in accordance with bulk PbSe behavior.[36] The same temperature dependence is suggested by the annealed 6% film emitting at longer wavelengths as temperature decreases, but we have not quantified the bandgap. We see that the as-grown 11% and 17% Ge film share the same atypical temperature dependence as PbSe. Upon annealing, however, we find that the positive temperature coefficient of the bandgap decreases in the 11% film, ultimately becoming negative in the 17% film. Now, the temperature dependence in the 17% Ge film is reversed, with the bandgap slightly widening with decreasing temperatures. This is seen as a subtle negative slope for 17% Ge. Nikolic also notes a narrowing of the bandgap with decreasing temperatures for PbSe, and a widening of the bandgap for bulk PbGeSe of 25% and 30% Ge composition.[14] Thus, we confirm that annealing PbGeSe produces epitaxial samples more representative of the bulk.

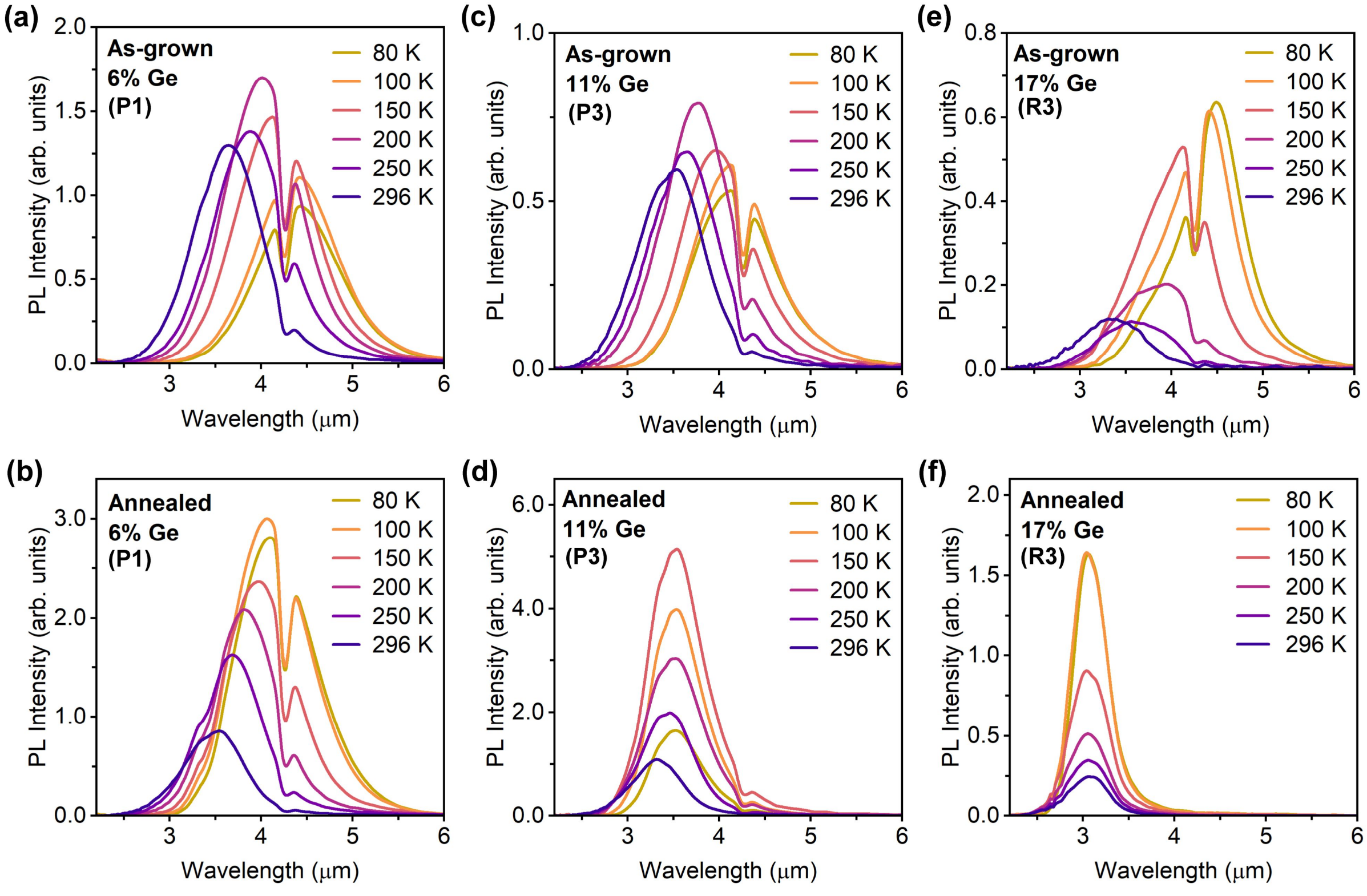

**Figure 5.** PL emission spectra at temperatures ranging from 80 K to 296 K for (a) as-grown 6% PbGeSe, (b) annealed 6% PbGeSe, (c) as-grown 11% PbGeSe, (d) annealed 11% PbGeSe, (e) as-grown 17% PbGeSe, and (f) annealed 17% PbGeSe. The compositions correspond to Samples P1, P3, and R3, respectively. The dip in the spectra at 4.3 μm is due to atmospheric $CO_2$ absorption.

Considering together our observations in the as-grown PbGeSe films of: (1) sluggish widening of the bandgap with Ge content, (2) severe broadening of X-ray peak, and (3) temperature trend of bandgap inconsistent with bulk PbGeSe, we posit these are signatures of a compositionally inhomogeneous microstructure. The PL emission in the as-grown PbGeSe films may arise predominantly from narrower bandgap PbSe-rich regions in the matrix of higher GeSe content, and thus the emission characteristics more closely resemble those of the PbSe with a positive temperature coefficient of the bandgap. The greatly reduced temperature sensitivity of the bandgap of PbGeSe is useful for uncooled non-dispersive infrared (NDIR) spectroscopy applications, where wavelength calibration due to temperature drift typically introduces overheads. The mechanism for negative temperature dependence of the bandgap in PbGeSe is not clear yet, but some authors point to Ge off-centering.[19] Additionally, with an eye towards carrier confinement in PbGeSe/PbSe heterostructures, the fortuitous opposite sign of bandgap change with temperature of PbGeSe and PbSe has the potential to provide additional confinement at low temperatures. We find that the bandgap of annealed 17% PbGeSe is about 110 meV greater than thin film PbSe at 200 K, and this difference grows to 160 meV at 80 K, for a confinement of approximately 0.4 meV/K.

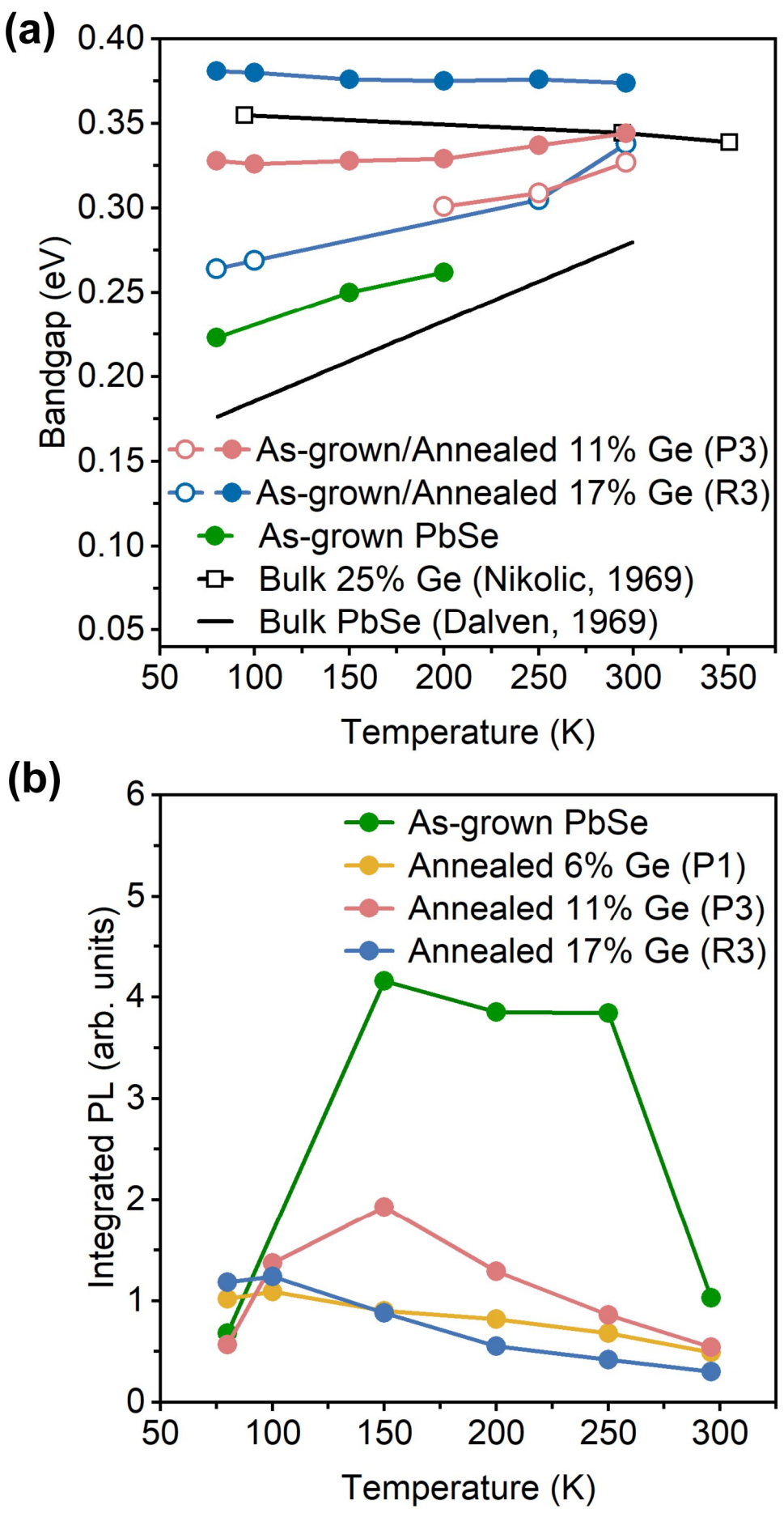

**Figure 6.** (a) Extracted bandgap vs. temperature for as-grown and annealed PbGeSe films with increasing compositions of 6%, 11%, and 17% (Samples P1, P3, and R3), including as-grown 195 °C PbSe film. Bulk 25% PbGeSe (Nikolic, Ref. 14) and bulk PbSe (Dalven, Ref. 36) are shown for reference. (e) Integrated PL signal vs. temperature for 195 °C PbSe and annealed PbGeSe with compositions of 6%, 11% and 17% (Samples P1, P3, and R3).

Figure 6b graphs the PL intensity at temperatures between 80–296 K for as-grown PbSe and annealed 6%, 11%, and 17% PbGeSe samples. The PbSe film shows a distinct reduction in PL signal for the lowest temperatures following our previous work conducted on this material for reasons not entirely clear to us.[37] The 11% Ge film exhibits a similar pattern in intensity, although less dramatic. In contrast, the PL signal for annealed 6% and 17% Ge film improves roughly 2−4× between 296 K and 80 K. In light of PL intensity trends in the 11% Ge film leaning towards PbSe, it is surprising to see trends in the 6% film lean closer towards 17% Ge behavior. This may be related to the addition of Ge and annealing conditions that change the dominant defect population in particular concentration regimes. Overall, these intensity trends are promising for PbGeSe/PbSe heterostructures and compare favorably to the incumbent PbEuSe, which sees a severe decrease in PL intensity at 100 K.[2,11]

## IV. CONCLUSIONS

We demonstrate a synthesis route for wider bandgap PbGeSe epitaxial films on GaAs, finding blue-shifted PL with increasing Ge content consistent with bulk crystals. Compared to PbSe, we achieve a 55 meV widening at room temperature with 17% Ge, and this increases to 160 meV at 80 K due to the opposite sign of the temperature coefficients of bandgap for PbSe and PbGeSe. The ultimate limits of Ge incorporation in the rocksalt phase by molecular beam epitaxy have not been explored in this work. Even within this range, adding Ge to PbSe compares favorably with Eu in PL intensity at cryogenic temperatures. Nevertheless, Ge does reduce PL intensity at room temperature and points to nonradiative centers introduced during synthesis or processing which need to be understood. The post-growth high temperature annealing method necessitated by poor GeSe sticking during growth of PbGeSe is not ideal for sharp composition profiles in heterostructures due to rapid intermixing. Our results with PbGeSe, however, show immediate potential to improve IV-VI devices such as detectors or optically pumped emitters in the SWIR-MWIR range where thick films are called for.

## ACKNOWLEDGMENTS

This work was supported by the Laboratory Directed Research and Development program at Sandia National Laboratories, a multimission laboratory managed and operated by National Technology and Engineering Solutions of Sandia LLC, a wholly owned subsidiary of Honeywell International Inc. for the U.S. Department of Energy's National Nuclear Security Administration under contract DE-NA0003525. This paper describes objective technical results and analysis. Any subjective views or opinions that might be expressed in the paper do not necessarily represent the views of the U.S. Department of Energy or the United States Government. The authors also gratefully acknowledge support via the NSF CAREER award under Grant No. DMR-2036520 for MBE operations. Characterization work was performed in part at the Stanford Nano Shared Facilities (SNSF), supported by the National Science Foundation under award ECCS-2026822.

## AUTHOR DECLARATION

### Conflict of Interest

The authors have no conflicts to disclose.

## DATA AVAILABILITY

The data that support the findings of this study are available from the corresponding author upon reasonable request.

**Supporting Information**

**Epitaxial PbGeSe thin films and their photoluminescence in the mid-wave infrared**

Kelly Xiao[1], Bryce Wong[1], Jarod Meyer[1], Leland Nordin[2,3], Kunal Mukherjee[1*]

[1]Department of Materials Science and Engineering, Stanford University, Stanford, CA 94305, USA

[2]Department of Materials Science and Engineering, University of Central Florida, Orlando, FL 32816, USA

[3]CREOL, The College of Optics and Photonics, University of Central Florida, Orlando, FL 32816, USA

*Corresponding author: kunalm@stanford.edu

## A. Temperature-Dependent Photoluminescence Spectra of Reference PbSe Film

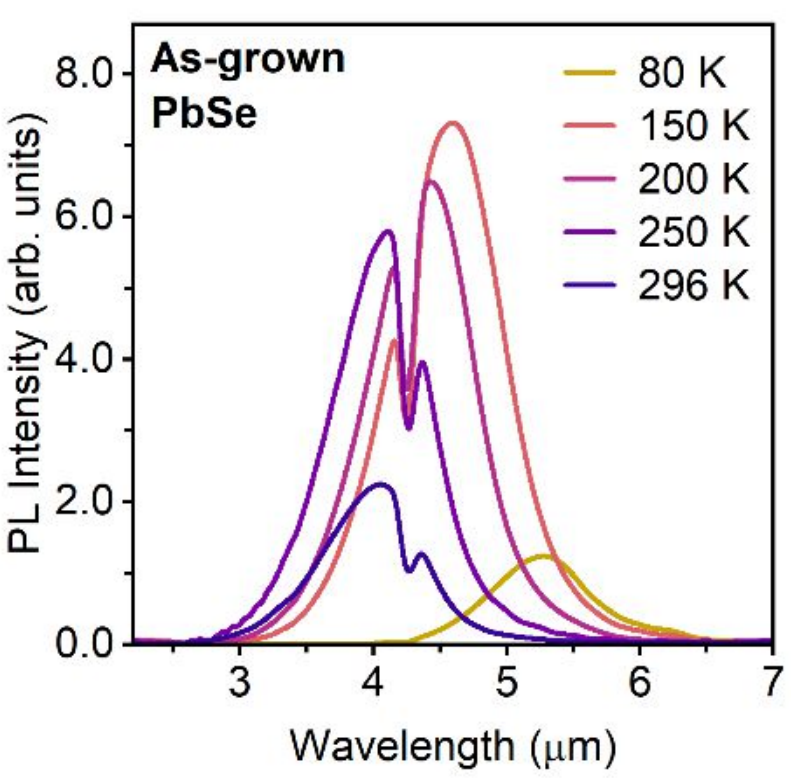

**Figure S1.** Photoluminescence spectra from 80–296 K for thin film PbSe grown at 195 °C. The dip in spectra at 4.3 microns arises from atmospheric $CO_2$ absorption.

Temperature-dependent photoluminescence measurements on PbSe show a redshift in the peak emission wavelength with decreasing temperature.